# Fake scientific journals are here to stay

Enrique Orduña-Malea

The iMetrics Lab. Departmento de Comunicación Audiovisual, Documentación e Historia del Arte, Universitat Politècnica de València, Valencia (España)

Email: enorma@upv.es

**Abstract:**

Scientific publishing is facing an alarming proliferation of fraudulent practices that threaten the integrity of research communication. The production and dissemination of fake research have become a profitable business, undermining trust in scientific journals and distorting the evaluation processes that depend on them. This brief piece examines the problem of fake journals through a three-level typology. The first level concerns predatory journals, which prioritise financial gain over scholarly quality by charging authors publication fees while providing superficial or fabricated peer review. The second level analyses hijacked journals, in which counterfeit websites impersonate legitimate titles to deceive authors into submitting and paying for publication. The third level addresses hacked journals, where legitimate platforms are compromised through cyberattacks or internal manipulation, enabling the distortion of review and publication processes. Together, these forms of misconduct expose deep vulnerabilities in the scientific communication ecosystem, exacerbated by the pressure to publish and the marketisation of research outputs. The manuscript concludes that combating these practices requires structural reforms in scientific evaluation and governance. Only by reducing the incentives that sustain the business of fraudulent publishing can the scholarly community restore credibility and ensure that scientific communication fulfils the essential purpose of reliable advancement of knowledge.

**Statements and Declarations**

*Funding:* Grant PID2022-142569NA-I00, funded by MCIN/AEI/ 10.13039/501100011033 and by "ERDF A way of making Europe"

*Competing interests:* The author has no competing interests to declare relevant to this article's content.

ChatGPT v5 was used to polish the English writing.

This preprint manuscript is an updated English version of the following post originally published in Spanish in Aula Magna:

>Orduña-Malea, E. (2025). Las revistas científicas falsas han llegado para quedarse. *Aula Magna 2.0* [blog]. https://doi.org/10.58079/15135

*The rise of scientific fraud*

The production of fake research has become a thriving business (**Abalkina et al.**, 2025). Harsh as it may sound, it is an undeniable reality. There is clear, manifest, and explicit evidence of a breakdown in the scientific communication ecosystem—an outcome of multiple, interconnected causes, perpetuated by various agents within the system, including authors, peer-review agencies, journals, and research institutions. In many cases this occurs unconsciously; in others, out of ignorance; and, in a few, deliberately.

The consequences are severe, not only for the scientific community but also for society at large, which learns, makes decisions, and builds values based on scientific knowledge. Today, scientific journals have largely ceased to act as guarantors of verified information and, in some cases, have even become yet another source of misinformation.

A quick search in the Scopus database for scientific publications that include terms[1] related to scientific fraud and academic misconduct in their titles or keywords reveals a total of 33,236 papers (as of 17 October 2025). Among these are a significant number of notes (3,458 publications), short reports (259), and editorials (187). Notably, a large proportion are unauthored (5,765 publications with no authorship data). These results illustrate a growing interest in the topic within the scientific community (**Figure 1**). Publication peaks coincide with some well-known cases, such as the Sokal affair and his scholarly hoax (1994), the Schön scandal at Bell Labs (2002), the Hwang scandal at Seoul National University (late 2005), and the Stapel scandal at Tilburg University (2011) due to scientific misconduct, and the Hindawi scandal on massive retractions (2023).

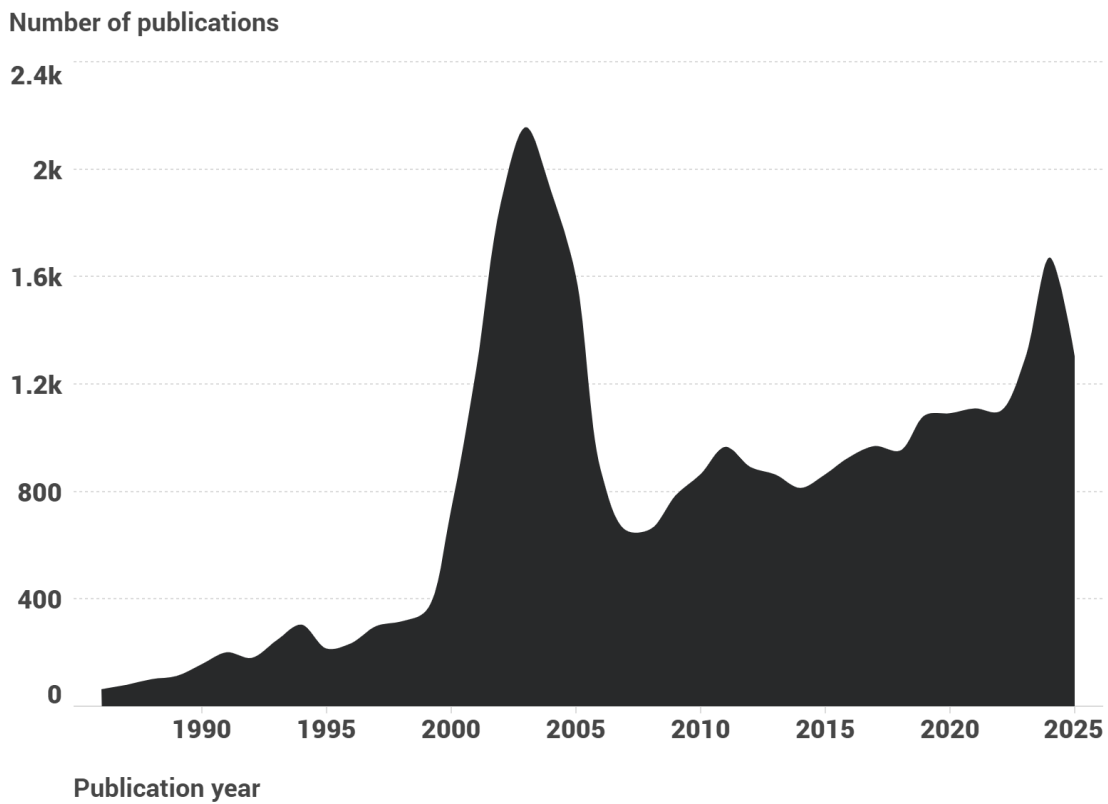

Figure 1. Evolution of scientific publications on scientific fraud
Source: Scopus. Note:

Fraud sometimes originates from the authors themselves, whether as an attempt to enhance their scientific reputation or to overcome specific promotion or evaluation processes. Examples of inappropriate practices include unjustified mass self-citation, excluded authorship (ghost authorship), false authorship (gift authorship), unsolicited authorship by prestigious individuals, plagiarism, data fabrication, data manipulation, duplication of publications, or salami publication, among others. Authors responsible for such works may also engage in malpractice when acting as reviewers, for instance by using coercive citations or issuing unjustified rejections or acceptances.

Generative Artificial Intelligence (GenAI) tools also have the potential to exacerbate these problems—for example, through the fabrication of bibliographic references (**Orduña-Malea and Cabezas-Clavijo**, 2023; **Walters and Wilder**, 2023) that can pass undetected through the supposed review process.

In other cases, fraud originates from the journals themselves, which is the focus of this article. In such cases, we can distinguish three main types: predatory, hijacked, and hacked journals, which will be discussed below. Specifically, this piece concentrates on hijacked and hacked journals.

*Level 1: Predatory journals*

These are publishers that approach scientific publishing primarily as a business, charging authors for each accepted and published article. Under this model, they seek to accept as many submissions as possible, levying a fee known as an Article Processing Charge (APC)—a euphemism for the cost of their "service." To attract more articles, such journals often offer rapid, superficial, and in some cases entirely fabricated or nonexistent peer reviews.

Some journals are established as predatory from the outset, while others transition from previously reputable and well-known titles following their acquisition by predatory publishers (**Martín-Martín and Delgado López-Cózar, 2025**). This "fast publication" system perfectly serves a segment of the scientific community that thrives despite mediocrity—often composed of researchers supported by public funds—who flourish within the system and occupy prominent positions. In this environment, prestige, reputation, and scientific impact are commodities, purchased as easily as one might buy a chair on Amazon: fast, convenient, and effective.

However, identifying and characterising predatory journals remains problematic (**Beall, 2015**), as the boundary between fraud and low quality is often blurred. It is important to clarify that the predatory nature of a journal does not necessarily arise from its open-access model, although open access has facilitated the shift from subscription-based to pay-per-publication business structures. The real concern lies in how the journal operates—whether open or closed—not in its access model per se.

*Level 2: Hijacked journals*

In this case, the journals in question generally operate correctly and may even enjoy a certain degree of prestige within the academic community. However, external actors clone the journal's website—creating counterfeit copies of the original with the intent to deceive the scientific community.

The first evidence of cloned or hijacked journals appeared around 2011 (**Jalalian and Dadkhah**, 2015). The phenomenon spread rapidly, affecting titles such as Archives des Sciences and Wulfenia, which at the time did not even have their own websites (**Butler**, 2013), or the International Development Planning Review, whose editors began receiving messages about publication fees despite the fact that the journal did not charge any (**Ryan**, 2024)—a clear sign that something was amiss.

The aim of journal hijacking is to lure researchers into submitting their manuscripts and paying for the publication service. To achieve this, fraudsters create websites identical to those of legitimate journals to appear credible. They often fill their tables of contents with fabricated or plagiarised articles from other sources to give the impression of an active publication.

The methods used by scammers vary widely. In some cases, they create a website for a journal with no existing web presence, making deception easier as there is no legitimate site for comparison. In other cases, they purchase a journal's web domain after it has expired or been abandoned. They may also register domains that closely resemble the original, making it very difficult to detect that the website is fraudulent.

For example, in 2024 the International Journal of Latin American Religions (https://link.springer.com/journal/41603) was cloned, with the fake version hosted on a similar-looking site (https://springer.nyc). Likewise, the Journal of Academic Ethics was cloned via a deceptive URL (http://jae.revistas-csic.com/index.php/jorunal), which closely mimics the domain of the CSIC journals portal (revistas.csic.es). It is somewhat ironic that a journal devoted to academic ethics should itself become the target of such fraud.

Cloned journals include the works submitted by authors in counterfeit issues and volumes to create the illusion of legitimate publication. To reinforce this, they assign fake or fabricated DOIs, or reuse those belonging to other publications, making it extremely difficult at times to recognise that one is navigating a fraudulent site.

The techniques employed are becoming increasingly sophisticated, and the fake websites are often highly convincing. In fact, these fraudulent sites sometimes achieve higher rankings on Google than the legitimate ones, thanks to the intensive use of search engine optimisation (SEO) techniques—as Clarivate Analytics itself has warned—thus reinforcing the illusion of credibility and visibility.

Although editorial teams from many hijacked publishers have contacted Google and Google Scholar to request that these fake websites be deindexed or demoted, they have received no response. This is remarkable given the enormous responsibility and influence these search engines hold as gateways to scientific information—particularly for younger researchers.

Those behind such fraudulent journals have achieved some notorious milestones in the history of academic deception. Some have successfully contacted major bibliographic databases such as Scopus, impersonating legitimate editors and persuading database managers to include links to the counterfeit websites in their journal directories. Once the databases became aware of the problem, they began removing these links (**Brainard**, 2023).

In a few cases, fraudulent journals even managed the unthinkable: articles published on their fake websites were indexed in bibliographic databases (**Abalkina**, 2024a). The fraud was sometimes uncovered when discrepancies emerged between the number of published articles and the journal's expected publication frequency.

It is important to note that articles published in fake journals are not necessarily fraudulent themselves. Some authors may have conducted sound, legitimate research but unknowingly submitted their manuscripts to counterfeit websites. However, studies have shown that many of these journals also contain numerous fraudulent or plagiarised papers (**Abalkina**, 2024b)—some fabricated to simulate activity, and others deliberately submitted by authors aware of the lack of peer-review controls and the possibility of being indexed in databases.

Given the growth of cloned journals and the risks they pose to the academic community, several directories have been developed to collect and disseminate information about hijacked titles. One early example is [Beall's list of hijacked journals](), now outdated. A key current resource is the [Retraction Watch Hijacked Journals Checker]()—a list compiled by international expert Anna Abalkina in collaboration with Retraction Watch—which included 380 journals as of 12 October 2025.

However, identifying such journals remains a laborious task. Many cloned journals take down their websites once detected, making follow-up difficult. In many cases, these websites are not even archived in the Internet Archive, which complicates the collection of evidence of their existence.

Some criteria for identifying cloned journals include:

1) *Check the journal's URL*, especially the main domain name, to verify that it corresponds to the legitimate publisher. Fraudulent journals often include links to the real site to appear credible, so any differences in the domain should raise suspicion.
2) *Check the DOI assigned to the article*. Click the DOI link to confirm that it redirects to the correct publication. Then verify its presence in Crossref and ensure that the link on screen matches the one displayed in the browser's address bar.
3) Check the receipt, acceptance, and publication dates. Extremely short timeframes may indicate fraud or inadequate peer review. While editors sometimes manipulate dates for reasons unrelated to quality, this remains a useful warning sign.
4) *Check the authenticity of the editorial team*. Fraudulent journals frequently fabricate editorial boards, either inventing names or—more convincingly—using the names of real scholars taken from other journals without their knowledge or consent.
5) *Check whether the journal charges publication fees*. Hijacked journals often impersonate open-access titles that normally do not charge authors. Their profit lies in collecting fees—typically ranging from about €500 to €2,000 or more—for supposed "publication services."

*Level 3: Hacked Journals*

A third category comprises hacked journals. In these cases, the issue is not impersonation but unauthorised access to the administrative area of a journal's website, allowing manipulation of the review or publication process, as well as other types of information published by the journal.

In many instances, such attacks do not specifically target a single journal but rather the broader organisation hosting it. In recent years, numerous universities and research institutions have suffered cyberattacks that compromised their entire web domains, including their journal portals. For example, the RCUB (the Scientific Journals Portal of the University of Barcelona) [officially reported an attack on its servers](). These attacks are often not aimed at interfering with editorial processes but rather at inserting links to fraudulent websites in order to boost their visibility and deceive search engines, which interpret incoming links from reputable domains as a sign of credibility (**Orduña-Malea**, 2021). In other cases, the objective is to steal data and demand ransom payments. In other situations, however, the attacks specifically target scientific journals—particularly platforms hosting multiple titles, which often manage numerous collections and are therefore complex to administer.

Sometimes, manipulation occurs without any formal cyberattack, but rather through the actions of authors and reviewers themselves, who act as Trojan horses. For instance, an author may invent a co-author—a detail that often goes unchecked. The well-known case of a researcher who for years listed his [cat as a fictitious co-author](#) or the [dog serving on the editorial board of medical journals](#) illustrates the absurdity of this loophole. When another researcher later submits a paper on a related topic, the journal might assign the fabricated co-author as a reviewer, meaning the original author effectively evaluates their own work.

When the number of fake reviewers within a journal becomes substantial, it can distort publication outcomes and enable the provision of "guaranteed" publishing services to paying clients desperate for acceptance. This, in turn, creates pyramid-like business models yielding enormous profits. Such fraud is especially difficult to detect, as it originates within the legitimate journal itself.

### *Final Remarks*

Eliminating—or at least mitigating—the effects of fake journals, particularly cloned ones, is an extremely complex challenge. One possible solution could be the development of digital certification systems to be implemented on the websites of journals that meet specific quality standards. Additionally, funding observatories to monitor journal quality and introducing punitive regulations to exclude those engaged in fraudulent activities from the scholarly ecosystem are measures worth considering.

However, in my view, the problem of fake publications and fraudulent journals (predatory, cloned, or hacked) will persist as long as authors feel compelled to publish—since this imperative fuels the entire business. Fraudulent organisations will cease their activities only when scientific publishing is no longer perceived as a sufficiently profitable enterprise. Achieving this requires a radical transformation of scientific evaluation and promotion processes.

Evaluation agencies and committees must acknowledge that mere co-authorship or citation does not equate to relevance, impact, or prestige. These can be obtained through personal networks—or even purchased in academic marketplaces. They must also recognise that the scientific community now includes individuals with little or no genuine scientific training—those unable to present their research coherently before experts, incapable of independently designing a study or writing a paper, yet able to construct an apparently strong academic record thanks to the current communication system cracks, and even to serve on evaluation committees themselves.

Faced with an ever-expanding community of imposter researchers—individuals more interested in publishing for money or career advancement than in genuine research—only a profound transformation of the scientific communication and evaluation system will bring an end to fraudulent journals.

**Notes**

[1] Search terms: Academic ethic*; Academic fraud; Academic integrity; Academic misconduct; Cloned journal*; Fake conference*; Fake journal*; Hijacked journal*; Paper mill*; Plagiarism; Predatory journal*; Predatory publish*; Publication ethic*; Questionable journal*; Scientific integrity; Scientific misconduct; Retraction Of Publication As Topic